\documentclass[twocolumn,epjc3-upd]{svjour3}  
\smartqed  
\RequirePackage{fix-cm}
\RequirePackage{graphicx}
\RequirePackage{subfigure}
\usepackage{ulem}
\usepackage{epsfig}  
\usepackage{multirow}
\usepackage{rotating}
\usepackage{subfigure}
\usepackage{xcolor}
\usepackage{xspace}
\usepackage{siunitx}
\usepackage{booktabs}
\usepackage[switch]{lineno} 
\journalname{Eur. Phys. J. C}
\DeclareGraphicsExtensions{.pdf,.png,.jpg,.mps,.eps}

\def\agcdm{$^{110{\rm m}}$Ag }
\def\agcdmn{$^{110{\rm m}}$Ag}
\def\agcd{$^{110}$Ag }

\def\cdcd{$^{110}$Cd }

\def\tect{$^{130}$Te }
\def\tectn{$^{130}$Te}

\def\xectn{$^{130}$Xe}

\def\pbdd{$^{210}$Pb }

\def\udt{$^{238}$U }
\def\udtn{$^{238}$U}
\def\thdt{$^{232}$Th }
\def\thdtn{$^{232}$Th}
\def\tld{$^{208}$Tl }
\def\tldn{$^{208}$Tl}
\def\bidq{$^{214}$Bi }

\def\rndd{$^{222}$Rn }
\def\rnddn{$^{222}$Rn}
\def\kq{$^{40}$K }
\def\kqn{$^{40}$K}
\def\coss{$^{60}$Co }
\def\cosn{$^{60}$Co}

\def\Qbb{$Q_{\beta\beta}$}
\def\BBz{$0\nu\beta\beta$~}
\def\BBzn{$0\nu\beta\beta$}

\def\BBd{$2\nu\beta\beta$~}

\def\ca{$\sim$}

\def\pom{$\pm$ }

\def\teod{TeO$_2$~}
\def\teodn{TeO$_2$}
\def\be{\begin{equation}}
\def\ee{\end{equation}}

\def\ciccio{5$\times$5$\times$5 cm$^3$ }

\def\partname{Section}
\def\figurename{Fig.}
\def\tablename{Table}
\def\qz{CUORE-0}

\begin{document}
\title{The projected background for the CUORE experiment}
\author{
C.~Alduino\thanksref{USC}
\and
K.~Alfonso\thanksref{UCLA}
\and
D.~R.~Artusa\thanksref{USC,LNGS}
\and
F.~T.~Avignone~III\thanksref{USC}
\and
O.~Azzolini\thanksref{INFNLegnaro}
\and
T.~I.~Banks\thanksref{BerkeleyPhys,LBNLNucSci}
\and
G.~Bari\thanksref{INFNBologna}
\and
J.W.~Beeman\thanksref{LBNLMatSci}
\and
F.~Bellini\thanksref{Roma,INFNRoma}
\and
G.~Benato\thanksref{BerkeleyPhys}
% ...
}

\institute{
Department of Physics and Astronomy, University of South Carolina, Columbia, SC 29208, USA\label{USC}
\and
Department of Physics and Astronomy, University of California, Los Angeles, CA 90095, USA\label{UCLA}
\and
INFN -- Laboratori Nazionali del Gran Sasso, Assergi (L'Aquila) I-67100, Italy\label{LNGS}
\and
INFN -- Laboratori Nazionali di Legnaro, Legnaro (Padova) I-35020, Italy\label{INFNLegnaro}
\and
Department of Physics, University of California, Berkeley, CA 94720, USA\label{BerkeleyPhys}
\and
Nuclear Science Division, Lawrence Berkeley National Laboratory, Berkeley, CA 94720, USA\label{LBNLNucSci}
\and
INFN -- Sezione di Bologna, Bologna I-40127, Italy\label{INFNBologna}
\and
Materials Science Division, Lawrence Berkeley National Laboratory, Berkeley, CA 94720, USA\label{LBNLMatSci}
\and
Dipartimento di Fisica, Sapienza Universit\`{a} di Roma, Roma I-00185, Italy\label{Roma}
\and
INFN -- Sezione di Roma, Roma I-00185, Italy\label{INFNRoma}
% ...
}

\thankstext{fn1}{Deceased}
\thankstext{fn2}{Presently at: INFN -- Laboratori Nazionali di Frascati, Frascati (Roma) I-00044, Italy}
\date{Received: date / Accepted: date}
\twocolumn
\maketitle

\begin{abstract}

The Cryogenic Underground Observatory for Rare Events (CUORE) is designed to search for neutrinoless double beta decay of \tect with an array of 988 \teod bolometers operating at temperatures around 10 mK. The experiment is currently being commissioned in Hall A of Laboratori Nazionali del Gran Sasso, Italy. The goal of CUORE is to reach a 90\% C.L. exclusion sensitivity on the \tect decay half-life of 9$\times$10$^{25}$ years after 5\,years of data taking. The main issue to be addressed to accomplish this aim is the rate of background events in the region of interest, which must not be higher than 10$^{-2}$\,counts/keV/kg/y. 
We developed a detailed Monte Carlo simulation, based on results from a campaign of material screening, radioassays, and bolometric measurements, to evaluate the expected background. This was used over the years to guide the construction strategies of the experiment and we use it here to project a background model for CUORE. 
In this paper we report the results of our study and our expectations for the background rate in the energy region where the peak signature of neutrinoless double beta decay of \tect is expected.
 
\end{abstract}

\section{Introduction}

CUORE is a ton scale experiment~\cite{Q_NIMA} with the primary physics goal of searching for neutrinoless double beta (\BBzn) decay of \tectn. Discovery of this phenomenon has been pursued now for several decades~\cite{DBDReview} and, if observed, it would provide crucial evidence for lepton number violation as well as open the door to physics models seeking to explain the matter-antimatter asymmetry in the universe.

Installed in the underground Hall A of Laboratori Nazionali del Gran Sasso (LNGS), Italy, CUORE is currently in the commissioning phase.
The experiment is the result of a long standing activity focused on the optimization of single particle thermal detectors (bolometers) based on \teod crystals \cite{MiDBD,rad2005,rad2006,qino-2011,qino-120Te,qino-exc,ccvr,Q-axions,Q-DM,Q0_PRL}.  
\teod-based bolometers have long been used in \BBz decay searches because their properties are well-matched to the requirements of such experiments; they have a very low heat capacity and exhibit extremely good energy resolution, while simultaneously serving as both the source of the \BBz decay and the detector. They also possess low intrinsic background \cite{ccvr,JCG312}, and can be operated stably for several years. Moreover the high natural isotopic abundance of the \BBz decay candidate \tect (34.17\%) \cite{isotopic_abundance}  avoids the needs for expensive isotopic enrichment at this stage.

MiDBD~\cite{MiDBD}, Cuoricino~\cite{qino-2011,qino-2008} and, more recently, \qz~\cite{Q0_PRL,Q0-InitialPerformances} marked important milestones in \teodn--based experiments, with successive improvements to \BBz decay sensitivity obtained through better energy resolution, increases in detector mass, and reduction of background.
CUORE is the latest step in this series of evolution: with respect to \qz, the active mass is 19 times higher (742\,kg of \teod or  206\,kg of \tectn) and the expected background event rate in the \BBz energy region is about 1/6th, \ca 10$^{-2}$\,counts/keV/kg/y. This corresponds to a 90\%\,C.L.  sensitivity on the \tect decay half-life of 9$\times$10$^{25}$ years in 5 years of exposure \cite{BanksKe}. 

The construction of CUORE required us to overcome a number of challenges, foremost among them was the design of an apparatus simultaneously meeting the stringent cryogenic and radioactivity constraints. This necessitated meticulous  material selection, optimization of production and handling protocols, development and validation of special cleaning procedures, and improvements in sensitivity of radioactive assay techniques. A detailed Monte Carlo simulation was used to evaluate the effects of contamination in the different elements of the apparatus. This  provided guidelines for radiopurity requirements and indicated where design modifications could mitigate effects from contamination that was unavoidable.
In this paper, we discuss the final steps of this work, presenting the \textit{background budget} of the CUORE experiment, namely the evaluation of the various contributions to the background event rate in the energy region of interest (ROI). 
\section{The CUORE experiment}\label{sec:exp}
\begin{figure}[!h]
\begin{center}
\includegraphics[width=1\linewidth]{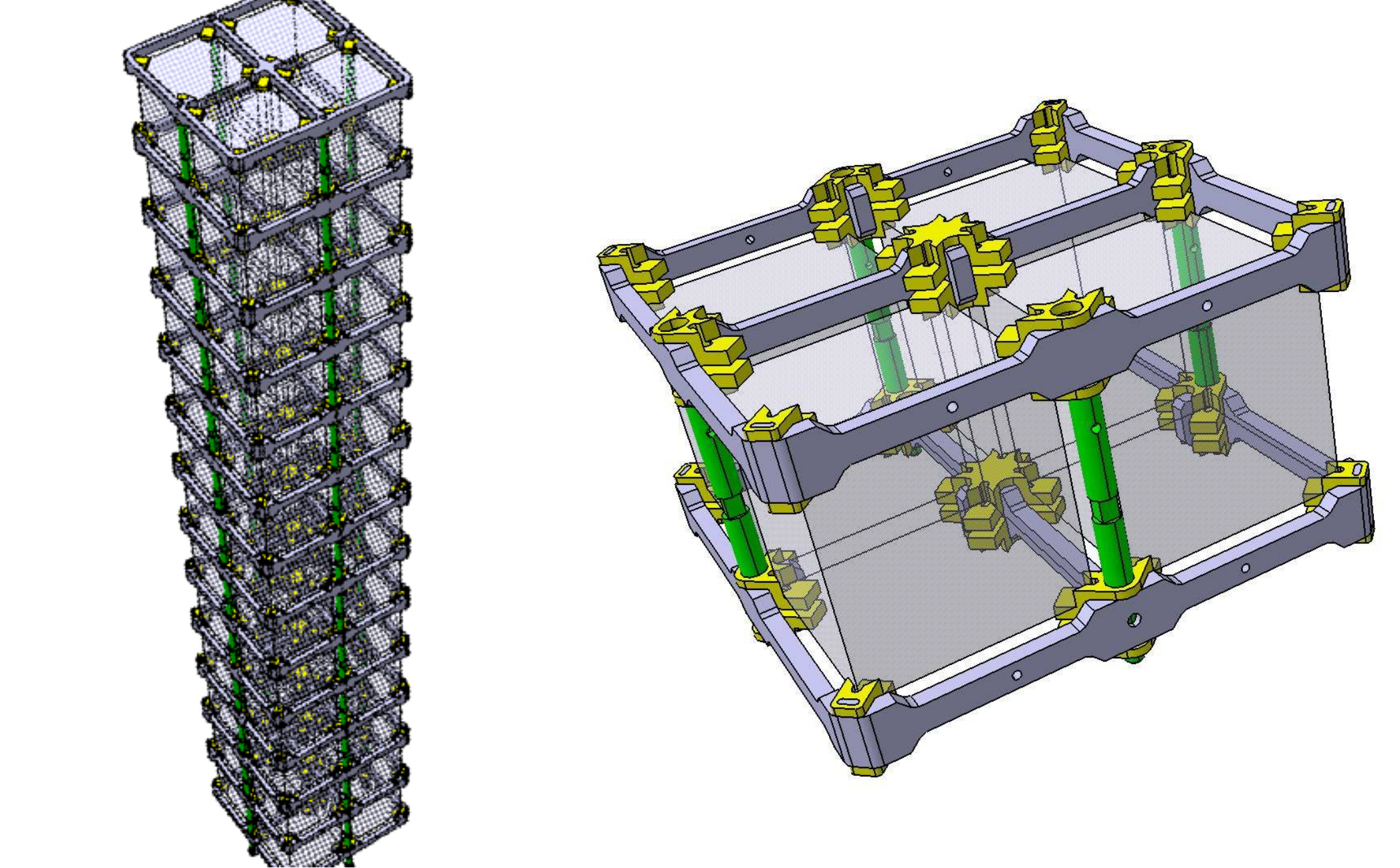}
\end{center}
\caption{Left: one of the 19 CUORE towers. Right: a 4-crystal plane. Two copper frames (grey) joined by four columns (green) form the mechanical structure that secures four crystals by means of polytetrafluoroethylen (PTFE) supports (yellow). }
\label{fig:torre}
\end{figure}

CUORE will search for the \BBz decay of \tect with a close--packed array of 988 \teod thermal detectors, operated at a temperature of \ca 10\,mK by means of a custom--made cryostat.
The CUORE detectors are arranged in a cylindrical matrix of 19 vertical towers, each one composed of 13 planes of 4 detectors modules supported by a copper frame. The rigidity of the structure is secured by four copper columns that connect each plane to the next one, as shown in \figurename~\ref{fig:torre}. All the copper components of the detector tower are made of NOSV copper, a special copper alloy suitable for cryogenic use produced by Aurubis~\cite{aurubis}.
Each detector is a 750\,g \teod cubic crystal ($5 \times 5 \times 5$\,cm$^3$) secured inside the copper frame by PTFE supports. These supports are designed to reduce the amount of mechanical stress on the crystals when cooled to cryogenic temperatures, and are the only mechanical parts in contact with the \teod crystals. A neutron-transmutation-doped (NTD) Ge thermistor \cite{95Haller} is glued with Araldit Rapid Epoxy onto the surface of each crystal to serve as a thermometer. The temperature increase produced by energy deposition due to particle interaction is converted into a voltage signal, which in turn is read out by a specially designed electronics chain. Electrical contacts to the detectors are obtained by means of Au bonding wires. These connect the thermistors to special flexible flat tapes (PEN-Cu cables~\cite{PENcalbes}) 
that run vertically along the tower, housed in NOSV copper wire trays. Both the Au bonding wires  and the PTFE supports act as thermal links between the crystal--thermometer system (i.e. the bolometer) and the heat bath (i.e. the dilution refrigerator), to restore the bolometers to the operating temperature  after each particle interaction. Finally, a small Si heater \cite{98SiHeater,12SiHeater} is glued onto each crystal and is used to generate reference thermal pulses for the off-line correction of temperature drifts. 

\begin{figure*}[htb]
\begin{center}
.\includegraphics[angle=-90, width=1\linewidth]{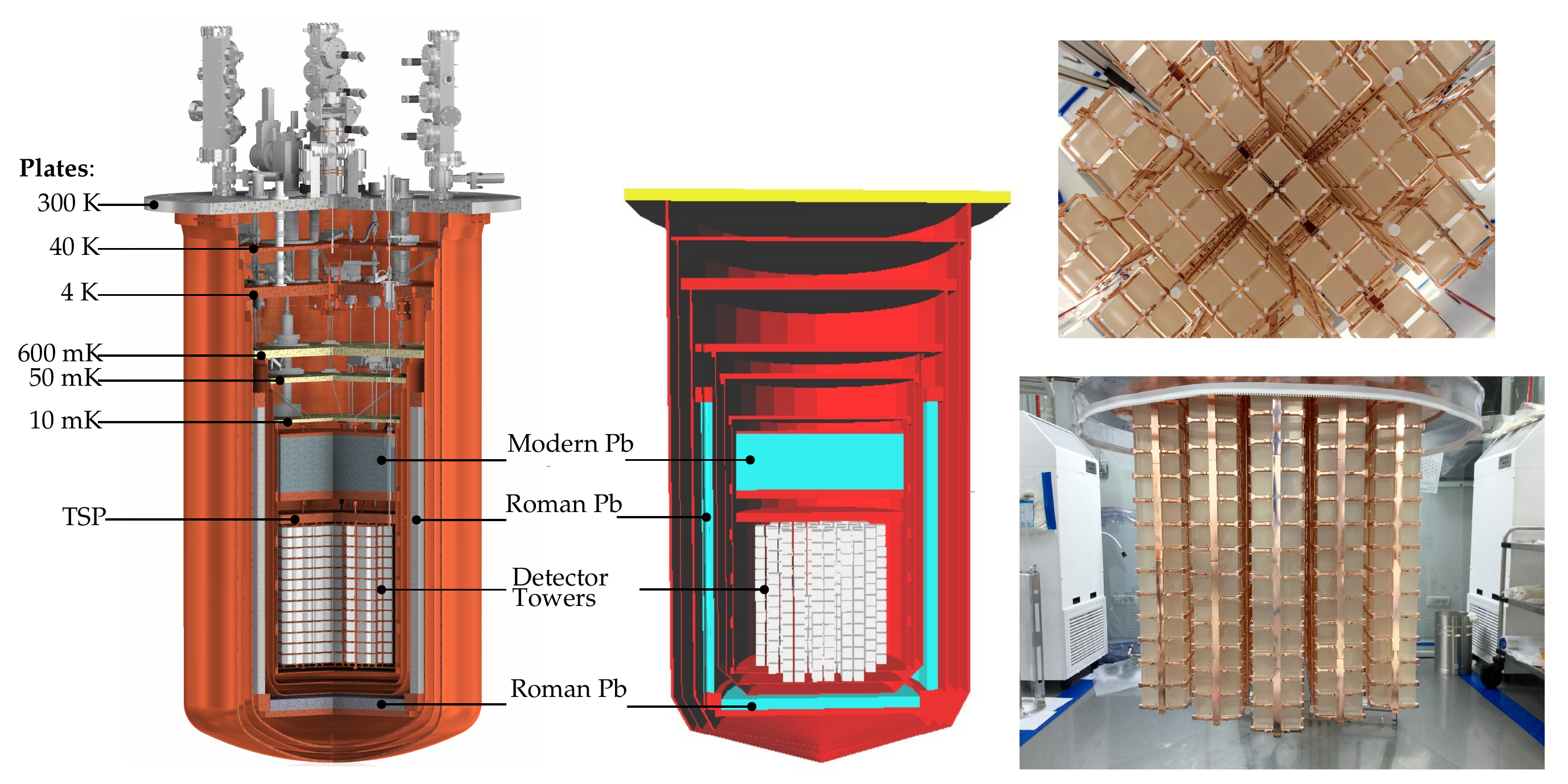}
\end{center}
\caption{Left: 3-dimensional view of the CUORE apparatus. Center: The CUORE setup as implemented in the Monte Carlo simulation. Cold lead shields are cyan-colored. Right: Pictures of the CUORE 19-tower array seen from the bottom and from the side, courtesy of Y. Suvorov.}
\label{fig:apparatus}
\end{figure*}

The CUORE detector array is operated in vacuum inside a custom-made cryogenic apparatus that complies with very stringent requirements regarding the lowest temperature reached, the mechanical vibration levels, the stability and reliability over long periods, and the radiopurity of all the materials in use. The complete system is shown in \figurename~\ref{fig:apparatus}. %
The cryostat is made up of six cylindrical nested copper vessels---acting as thermal shields---which are thermally anchored to the different temperature stages of the cryogenic system and labeled according to their working temperature: 300\,K, 40\,K, 4\,K, 600\,mK, 50\,mK, and 10\,mK thermal shields. Each vessel is closed on the top by a thick plate; all plates are made of copper but the 300\,K one, that is made out of stainless steel\footnote{The 300~K vessel is not shown in \figurename~\ref{fig:apparatus}. Its upper part (about 1/6 of its length) is also made of stainless steel.}. 
All the copper vessels and covers but the 10\,mK one are made of OFE copper (C10100 type), an oxygen-free copper alloy  suitable for e-beam welding and high-vacuum sealing.
The 300\,K---Outer Vacuum Chamber (OVC)---and the 4\,K---Inner Vacuum Chamber (IVC)---thermal shields are vacuum--tight. Also, the 40\,K and the 4\,K vessels are surrounded by several layers of super-insulation foil for cryogenic needs (not shown in \figurename~\ref{fig:apparatus}). 
The apparatus is a cryogen free cryostat, and the various temperature stages are maintained by a complex cooling system which includes a set of pulse tube cryocoolers and a custom made 
high-power  $^3$He/$^4$He dilution refrigerator. 

The detector towers hang from a thick copper disk---the Tower Support Plate (TSP)---placed inside and thermalized to the 10\,mK thermal shield. 
The TSP is mechanically decoupled from the rest of the experimental setup in order to minimize the transmission of mechanical vibrations to the detectors. Both the TSP and the 10\,mK shield are made of NOSV copper. Moreover they are covered, on the side facing the detector array, by thin NOSV copper tiles. This design allowed all the copper parts that face the detector, including the tower skeleton, to be cleaned with the same procedure specially tuned to minimize surface radioactive contaminants (which was not possible for the large and massive parts such as the thermal shield or the TSP).

Two cold lead shields are used to protect the array from the background contributions coming from the cryogenic apparatus. One is a 30\,cm thick disk of modern lead together with 6.4~cm of NOSV copper placed just above the TSP inside the 10\,mK shield, and thermally anchored to the 50\,mK shield, to protect the detector from radioactivity coming from above. The other is a 6\,cm layer of ancient Roman lead, thermally anchored to the 4\,K vessel, which shields the detector on the side and on the bottom.
The Roman lead is mechanically supported by stainless steel bars. Both the modern and Roman lead shields are thermalized by NOSV copper spacers.
Outside the cryostat, at room temperature, two shields surround the setup for the abatement of environmental neutron and $\gamma$ fluxes: a lead shield with \ca 25~cm minimal thickness to absorb environmental $\gamma$ rays, and an outer neutron shield, consisting of 18\,cm of polyethylene (to thermalize neutrons) and 2\,cm of H$_3$BO$_3$ powder (to capture thermal neutrons). 
%

For the periodic energy calibration of the detector, twelve thoriated tungsten wires are deployed into the cryostat; six are guided into NOSV copper tubes that are placed in fixed positions within the 10\,mK volume, and six are guided to the outside of the 50~mK shield. This allows for a near-uniform irradiation of all the crystals with $\gamma$ rays from the \thdt decay chain \cite{17Jeremy}.
\section{CUORE background sources}\label{sec:sources}
If it occurs in nature, \BBz decay can be detected by measuring the energy of the two emitted electrons. Since bolometers are calorimetric detectors, the events in which both electrons are stopped inside the emitting crystal produce a monochromatic peak in the energy spectrum. Particularly, the process \tect $\rightarrow$ \xectn\,$+2\beta^-$ is expected to produce (when both electrons are stopped inside the emitting crystal)  a line 
at \Qbb\,=\,2527.518\pom0.013\,keV, the decay transition energy~\cite{QValueFrank,QValueNick,QValueRaman}.

We define the \tect \BBz decay ROI as a 100~keV--wide region in the spectrum centered at \Qbb\ (2470--2570 keV). 
A number of sources in addition to \BBz decay can produce events in this energy region, including natural and artificial environmental radioactivity, cosmogenically activated isotopes, and cosmic rays. 
In most cases these sources produce a continuum in the ROI, that is explicitly chosen to exclude both the 2448 keV line of \bidq and the 2587 keV Te X-ray escape peak due to the \tld line at $\sim$2615~keV. The only peak that, given its proximity to \Qbb, cannot be excluded from the ROI is the \coss sum peak at 2505 keV, due to simultaneous detection in a single crystal of the two \coss $\gamma$ rays.

The only distinctive signature expected of a \BBz decay event in CUORE is the energy of the signal and its \textit{single-hit} characteristic. In the assumption of a uniform distribution of $\beta\beta$ decaying nuclei in the crystal, the probability  that both emitted electrons are contained inside the single crystal is \ca 88.4\%~\cite{Q0-analysis}. Therefore, in most cases a \BBzn\ decay
event results in a signal in only one crystal at a time (\textit{single-hit} event). On the contrary a large fraction of background events occur in more than one detector at a time (\textit{multi-hit} event). 
The rejection of \textit{multi-hit} events improves the signal-to-noise ratio in the ROI by reducing the background by a larger amount than the signal, and is therefore a powerful active background rejection tool in the CUORE experiment.

The most common background sources for experiments located in underground laboratories are the radioactive contaminants of the construction materials ~\cite{bkg_EXO,bkg_EDELWEISS,bkg_GERDA,bkg_KZ,bkg_MAJORANA}, namely:
\begin{itemize}
\item long--lived radioactive nuclei such as \kqn, \udtn, and \thdt (the latter two are progenitors of  radioactive decay chains);
\item anthropogenic radioactive isotopes (mainly produced by industrial processes or by human--induced nuclear activity in the atmosphere) such as \cosn, $^{137}$Cs, and $^{134}$Cs;
\item cosmogenically--produced radioactive isotopes (e.g. {\cosn} caused by fast nucleon interactions in copper and tellurium);
\end{itemize}
\noindent Another common source of background is the environmental flux of neutrons, muons and $\gamma$ rays at the experimental site. 
In the case of CUORE, the LNGS average rock overburden of $\sim 3600$\,m.w.e. strongly reduces the muon flux~\cite{CuoreExternal}. Moreover, the simultaneous effects of the outer neutron and lead shields, as well as the off-line rejection of \textit{multi--hit} events, are expected to reduce the background contribution of this source in the ROI to a negligible level, compared to background from radioactive contaminants, as discussed in~\cite{MuFlux1,MuFlux2,MuFlux3}. 
Therefore, in this paper we focus on radioactive contaminants, since they drive the sensitivity of CUORE.
A similar condition held also for CUORE precursors. In particular, three dominant sources were identified as contributors to the event rate recorded in the ROI of the MiDBD and Cuoricino experiments~\cite{qino-2008,EPJ2}:

\begin{enumerate}
\item multi-Compton events from the 2615~keV $\gamma$ rays of \tldn\footnote{This is the only environmental gamma line with a B.R.~$>$~1\,\% and an energy higher than \Qbb.} originating from \thdt  contamination of the cryostat;
\item  contamination by \udtn, \thdtn, and their decay products (including the surface implantation of \pbdd from environmental $^{222}$Rn),  on the surface of the \teod crystals;
\item  contamination by \udtn, \thdtn, and their decay products, on the surfaces of inert materials in close proximity to the \teod crystals, most likely the copper holder structure. 
\end{enumerate}

This result drove most of the strategies adopted for the realization of CUORE~\cite{Q0-detector} including: the minimization of both the amount of material used in the support structure and the space between the crystals (the latter improves the efficiency of background rejection through the \textit{ multi-hit} cut); the stringent control of material contamination; and the design of highly efficient shielding from radioactivity.
A completely new design was developed to achieve at least a 20-fold reduction of the $\gamma$ ray induced background (source 1 in the previous list) which dominated in the previous experiments.
Special cleaning processes were developed to reduce sources 2 and 3.
In particular all the \teod crystals were processed following a strict radiopurity protocol \cite{JCG312} with a final surface polishing which  reduced \thdtn, \udtn, and \pbdd surface contaminants $\sim$6 fold relative to what previously achieved. Copper is by far the most abundant material in the ambient surfaces directly facing the detectors. Therefore, a special cleaning protocol was optimized for all NOSV copper parts constituting the detector holder (frames, columns, wire trays, etc.) as well as for the thin copper plates which surround the whole detector array. The procedure was developed at Laboratori Nazionali di Legnaro, Italy, and includes a sequence of different steps of surface treatments---Tumbling, Electropolishing, Chemical etching, and Magnetron plasma cleaning. This particular methodology, called TECM cleaning, was shown to be the most effective method among those investigated for CUORE \cite{TTT}.  In order to avoid surface recontamination due to airborne radioactivity, which is primarily responsible for \rnddn contamination, particular care was devoted to the production, storage and assembly of all the detector components. These were handled and stored in nitrogen atmosphere and the 19 towers, once assembled, were continuously flushed with nitrogen. The final installation of the CUORE towers in the cryogenic system took place in a dedicated clean-room environment, with continuously circulating radon-depleted atmosphere and constant monitor of the radon level.  Details can be found in~\cite{JCG312,14AssemblyLine,LNGS15}.

To test the effectiveness of these background reduction strategies, the first tower of \teod detectors constructed in the CUORE assembly line \cite{14AssemblyLine} was installed in the same cryostat used for the MiDBD and Cuoricino. 
This tower, named \qz~\cite{Q0-detector}, was also operated in Hall A of LNGS serving both as a technical prototype for CUORE as well as a stand-alone \BBz decay experiment~\cite{Q0_PRL}. Thanks to the high quality of its data, a very detailed background reconstruction was possible, disentangling the major sources contributing to the ROI event rate. While no change was expected in the background contribution of the cryostat shields  because the new tower was operated in the older cryostat, a reduction of the contribution of the other two sources was expected from the improved background abatement protocols. \qz\ results proved to be in good agreement with expectations~\cite{Q0_2nu}, as shown in Table~\ref{tab:Cuore0ROI} (see also Fig.\,15 in~\cite{Q0_2nu}). 
The observed background was reduced $\sim$2.6-fold relative to Cuoricino demonstrating the effectiveness of the cleaning protocols.
\begin{table}
\setlength{\tabcolsep}{3pt}
\begin{center}
\caption{List of the background sources contributing to the \BBz decay
ROI of the \qz\ experiment (from~\cite{Q0_2nu}). The continuum event rate in this region (i.e. excluding the \coss sum peak) is $0.058\pm0.006$~counts/keV/kg/y)~\cite{Q0-analysis}. Column (2) reports the fractional contribution of the different sources, as obtained by the \qz\ background reconstruction. The \textit{Holder} is the structure that held the crystals, made from \textit {CuNOSV} and reproduced identically in CUORE. The \textit {Shields} represent both the cryostat external and internal shields used in \qz\ to shield the detector from environmental and setup radioactivity (see~\cite{Q0_2nu} for more details).}
\label{tab:Cuore0ROI} 
\begin{tabular}{lrcl}
\toprule
Component	&	\multicolumn{3}{c}{Fraction [\%]}	\\
\midrule
Shields					&74.4	 &$\pm$ &1.3	 \\
Holder					&21.4	&$\pm$ &0.7	 \\
Crystals 				&2.64	&$\pm$ &0.14		 \\
Muons					&1.51	&$\pm$ &0.06	 \\
\bottomrule

\end{tabular}
\end{center}
\end{table}

\section{CUORE Monte Carlo code}\label{sec:mc}
Our simulation is based on the 
{\sc GEANT4}~\cite{Agostinelli2003250} package; its architecture and its implementation of the physics are identical to the one adopted in the simulation successfully used for CUORE-0 background reconstruction  ~\cite{Q0_2nu}.

The simulation code generates and propagates primary and secondary
particles through the CUORE geometry until they
are detected in the \teod crystals. It outputs
the energy and time of the energy depositions; the time is used to properly take into account correlations in nuclear decay chains.
\begin{table*}[!htb]
\begin{center}
\caption{Elements of the CUORE experimental setup implemented in the Monte Carlo simulation, with the values of their masses and surface areas. Elements are grouped into two regions: the near region and the far one. Under the OFE category we included also a 3\% of NOSV copper parts located in the far region. The near region includes components close to the \teod detectors. Surface contamination is simulated only for elements in the near region, hence surface areas are listed only for them. The surface area indicated for NOSV copper parts only takes into account surfaces directly facing the detectors. The rightmost column lists the short name of the components used in this paper.}
\label{tab:masses}
\begin{tabular}{@{}llccl@{}}
\noalign{\smallskip}\toprule
Region & Element						&Mass [g]				&Surface [cm$^{2}$]		&Short name			\\
\noalign{\smallskip}\midrule
\noalign{\smallskip}
\multirow{7}{*}{Near} 
& \teod crystals						&7.42$\times$10$^{5}$	&	148200				&\textit{\teod}	\\
&Si heaters								&6.8					&	158					&\textit{heaters}	\\
&NTD Ge thermistors						&42						&	288					&\textit{NTDs}\\
&PEN-Cu cables							&389					&	1200				&\textit{PEN}\\
& Copper wire pads				     	&57		                &1140		            &\textit{pads}\\
&PTFE supports							&5500					&	29800				&\textit{PTFE}\\
&NOSV copper parts						&9.4$\times$10$^{5}$	&	278900			    &\textit{CuNOSV}\\
\midrule
\multirow{6}{*}{Far} 
&OFE copper parts		&7.0$\times$10$^{6}$	&-	&\textit{CuOFE}\\
&Superinsulation layers					&17$\times$10$^{3}$		&-	&\textit{SI}	\\
&Roman lead shield						&4.5$\times$10$^{6}$	&-					    &\textit{RomanPb}\\
&Modern lead shield						&2.1$\times$10$^{6}$	&-					    &\textit{ModernPb}\\
&Stainless steel rods					&15.2$\times$10$^{3}$	&-	&\textit{Rods}\\
&300K steel flange						&1.9$\times$10$^{6}$	&-	 &\textit{300KFlan}	\\
\noalign{\smallskip}\bottomrule
\end{tabular}
\end{center}
\end{table*}

We then take the output from {\sc GEANT4} and apply a detector response function and other readout features.

Our code is based on the  4.9.6.p03 version of the {\sc GEANT4} simulation toolkit, and employees the Livermore Physics List. It includes the propagation of photons, electrons, $\alpha$ particles and heavy ions (nuclear recoils from $\alpha$ emission), as well as neutrons and muons. All primary particles and secondaries are propagated down to keV energies, with a tracking cut optimization inserted in the different detector volumes to balance simulation accuracy and speed (e.g. the tracking cut in lead is set to 1~cm, while in copper to 1~mm).
The generation of nuclear transitions is based on an \textit{ad hoc} implementation of the G4RadioactiveDecay database, performing a concatenation of the tabulated single isotopes decays
to correctly simulate chains of radioactive decays in secular equilibrium taking into account their specific time structures. 


In the simulation the geometries of the detector, of the cryostat, and of the internal and external shields are reproduced (see \figurename~\ref{fig:apparatus}), 
namely: the \teod crystals, the copper structure holding the array (i.e. frames and columns), the PTFE supports, the wire trays, the NTD Ge thermistors, the calibration source guiding tubes, the lead support steel bars, the various thermal shields and other cryostat parts, the internal and external lead shields, and the external polyethylene shield.
In \tablename~\ref{tab:masses} we list the masses, surface areas, and materials corresponding to each element implemented in the Monte Carlo simulation. \textit{CuNOSV}, henceforth, includes all the NOSV copper components of the detector, i.e. frames, columns, wire--trays, calibration source tubes,  10\,mK shield, TSP, all the thin plates covering the TSP and the 10\,mK shield, and the two NOSV copper disks enclosing the internal modern lead shield (see \partname~\ref{sec:exp}). Similarly, \textit{CuOFE} indicates all the OFE copper parts of the experimental setup (mainly the cryostat thermal shields---see \partname~\ref{sec:exp}). 

In our convention, elements are geometrical volumes made of the same material that have a similar production history and therefore similar contamination levels. Each element of the simulation can be studied as an active source whose bulk and/or surface radioactive pollutants  are simulated independently in order to evaluate the effect of its contamination, particularly in the ROI. We group elements into two regions (\textit{near} and \textit{far}) according to their proximity to the bolometers. This is useful when discussing surface contaminations, as it will become clear in the following. 

The distribution of impurities in the bulk of the different materials is assumed to be spatially uniform, while impurities on the surfaces are modeled according to the diffusion process with an exponential density profile

\begin{equation}
\label{eq:exp}
	\rho=\rho_0 \times \exp(-x/\lambda),
\end{equation}

 \noindent where $\rho_{0}$ is the impurity density, $x$ is the distance from the surface into the bulk, and $\lambda$ is the mean penetration depth of the impurity. 
 
For reproducing the experimental data from the Monte Carlo simulation, we also model the detector time and energy response.
Assuming a detector response  similar to that of \qz\ detectors \cite{Q0-analysis}, we account for particle pulses (i.e. pulses following energy deposits in the crystals) with 
rise times of $\sim 0.05$\,s and two decay time components, a fast decay
time of  $\sim 0.2$\,s and a slower one of  $\sim 1.5$\,s. 
We account for the timing resolution of each crystal by summing energy depositions that occur in the same crystal within a time window of $\pm$ 5\,ms (pulses with longer time distances can be distinguished).
Once the simulated events are correctly correlated in time, the resulting energy depositions are smeared with a Gaussian energy response function assuming an average energy resolution of 5~keV FWHM (i.e. the design energy resolution at \Qbb\ of CUORE and realized in \qz\footnote{In the \qz\ experiment, the projected energy resolution was $5.1\pm0.3$\,keV FWHM at $Q_{\beta\beta}$.}\cite{Q0-analysis}).
Consistent with what is done in the experiment, we define coincident events as those occurring in multiple bolometers within a $\pm$ 5\,ms window. A multiplicity parameter is added to each event, given by the number of bolometers  in coincidence for that event. Single-hit events are identified by multiplicity\,=\,1. Finally, as done on \qz\ real data \cite{Q0-analysis}, a pile-up cut of 7.1~s around each event is implemented, rejecting pulses occurring in the same bolometer within 3.1~s before or 4~s after the selected event, and dead--time is accounted for.

\section{Material assay}\label{sec:simu-input}

An extensive screening campaign has preceded the selection and procurement of all materials used in the construction of CUORE (for details on material assay techniques commonly adopted in this field see~\cite{bkgLTR}). The focus was on the most ubiquitous natural contaminants \thdt and \udt (with their respective progenies), and on a few cosmogenically activated contaminants: \cosn, \agcd and \agcdmn~\cite{EttoreCosmogenic}. All these isotopes emit, in their decay chains, particles that have enough energy to mimic a \BBz decay. Even if the mentioned isotopes do not include all the natural or comogenic radionuclei, all other isotopes are expected to yield minor, if not negligible, contributions in the ROI. 

The selection of all the CUORE components comprised the certification of the material activity,
both for bulk and surface contaminations. Surface contamination can occur during the processes of machining and cleaning, or during exposure to contaminated air.  In addition, the effect of breaks in the secular equilibrium of a radioactive chain can be quite relevant for surface contaminations, as it is the case for \pbdd in the \udt chain. Indeed, \rndd emanation from any material containing \udt impurities produces excess concentrations of \pbdd (the only long-lived isotope in the \rndd progenies) in the air and dust that, in turn, can contaminate exposed components. 

The techniques adopted in the various phases of material screening include $\gamma$ spectroscopy with heavily-shielded High Purity Germanium (HPGe) diodes, to investigate bulk contaminations, and $\alpha$ spectroscopy with large-area low-background Silicon Surface Barrier (SiSB) diodes, to screen for surface contaminations. Additionaly, Inductively Coupled Plasma Mass Spectrometry (ICPMS) and Neutron Activation Analysis (NAA), both particularly suitable for small samples, were also used. Alpha and gamma spectroscopy and ICPMS analysis were performed at LNGS~\cite{04Laubenstein,Heusser-gempi}, Milano-Bicocca~\cite{13Sala}, Baradello Laboratory~\cite{04Brofferio} and Lawrence Berkeley National Laboratory (LBNL)~\cite{13Thomas}. Neutron activation analysis was carried out in collaboration with the Laboratory of Applied Nuclear Energy (LENA) in Pavia, Italy~\cite{NAA1,NAA2}.

All materials constituting the CUORE experimental setup, as well as those in contact with the detector components during the production and cleaning phases, have thus been carefully selected  according to their bulk contamination levels.
Then, for materials in the far region (see \tablename~\ref{tab:masses}), careful cleaning and storage with standard techniques are enough to ensure that surface contributions are negligible. This was routinely confirmed with SiSB detectors.
On the other hand,  in the near region the radiopurity requirement for material surfaces can be as strict as a few nBq/cm$^2$. In fact, in elements sufficiently close to the bolometers $\alpha$ and $\beta$ particles can also mimic a \BBz event. For these particles the detection efficiency is much higher (since the absorption probability is nearly 1), and even contaminants with a small activity can produce a relevant background rate in this case. In critical cases when the required sensitivity was not achievable with SiSB detectors,  NAA and ICPMS have been exploited to certify material surface contamination.

Finally, in the few cases where all the above techniques failed in reaching the required sensitivity, bulk and/or surface contamination levels were determined through particle spectroscopy with \teod detectors~\cite{ccvr,EPJ2,EPJ1}. These measurements were carried out underground at LNGS using arrays of CUORE-like \teod bolometers. One of the most remarkable results obtained in this case is the evaluation of the radioactive sources contaminating the \qz\ detector with an unprecedented precision, as will be discussed later. 

In the next sections 
the evaluation of contaminant activities in all the materials used for the CUORE construction will be presented. 
The evaluation of bulk contaminants will be discussed first, followed by surface ones. 
The \qz\ results, both for bulk and surface contaminations, will instead be discussed in two dedicated sections.

\subsection{Cosmogenic activation}\label{sec:activation}

Cosmogenic activation is a well-known mechanism for the production of radioactive nuclei in the bulk of materials, mainly through spallation processes. At sea level, cosmic rays are comprised of charged pions, protons, electrons, neutrons, and muons, with relative flux intensities of roughly 1:13:340:480:1420 \cite{RadioProtection}. Neutrons are clearly the dominant source of cosmogenic activation above ground. At the CUORE experimental site at LNGS, the cosmic ray flux is decreased to an almost negligible level (only muons survive the rock overburden), and hence cosmogenic activation is drastically reduced. Therefore, in this paper we analyze only activation of materials before their  underground storage.

In CUORE, the most abundant materials are \teodn, copper, and lead. Among these, only Te and Cu isotopes exhibit large enough cross sections for radioisotope production via cosmogenic activation.
In order to contribute to the ROI, cosmogenic nuclei must have a transition energy greater than \Qbb, a sizable production cross section, and a relatively long half-life (compared to the time scale of the experiment). Based on the neutron flux at sea-level \cite{04Gordon,cosmicflux}, the exposure time of copper and \teod to cosmic rays, and the measured or calculated neutron spallation cross sections \cite{act1,act2,act3,act4,act5,act6,act7,act8,BarbaraCosmogenic,EttoreCosmogenic}, two isotopes fulfill these criteria: {\cosn} and {\agcdmn}.

{\cosn} ($\tau_{1/2} = 5.27$ years, {\it Q} = 2.82~MeV) is produced both in copper and tellurium. It $\beta$-decays with the simultaneous emission of two $\gamma$ rays (1.17\,MeV and 1.33\,MeV). To mimic the energy of a \BBz event, both the $\gamma$ rays and the electron must deposit their energy in the same crystal, requiring \coss to be either in the \teod crystals or in the copper parts close to the bolometers (\textit{CuNOSV}). 

\agcdm ($\tau_{1/2} = 250$ days) is produced in tellurium. 
It can either $\beta$-decay to the stable isotope \cdcd (B.R. = $98.7\,\%$, {\it Q} = 3.01\,MeV), or it can make an isomeric transition to \agcd (B.R. = $1.3\%$, {\it Q} = 0.118\,MeV), which then further $\beta$-decays to \cdcd (B.R. = $99.7\,\%$ and {\it Q} = 2.89\,MeV). \agcd has a short half-life ($\tau_{1/2} = 24.6$ s) and will therefore be in secular equilibrium with \agcdm during the expected 5 years of CUORE data-taking.
Both \agcdm and \agcd emit numerous photons in their decays. The decay of either isotope can mimic a \BBz event when the total energy 
deposited in a single crystal by the photons and electrons is  within the ROI.  

To minimize the activation levels of \teod and copper (both NOSV and OFE), their exposure to cosmic rays was reduced as much as possible. Approximately 3 months elapsed from crystal growth to underground storage at LNGS for the \teod crystals, followed by an average cooling time of 4 years before use. 4 months elapsed from casting of the raw materials to underground storage for all the machined and cleaned {\it CuNOSV} components. Requirements were less stringent for OFE copper since this is used in the far region.
The activation levels of \teod crystals at the start of CUORE have been estimated based on the results of neutron and proton activation measurements published in \cite{BarbaraCosmogenic} and  \cite{EttoreCosmogenic}. Estimation of activation levels for {\it CuNOSV} follows \cite{SusanaCosmogenic} and incorporates the \qz\ results. 
The estimates of the activation levels are given as conservative upper limits, evaluated approximately one year prior to the beginning of CUORE data taking.
These are:
\begin{itemize}
\item $<20$\,nBq/kg of \agcdm + \agcd (in secular equilibrium with each other) in the {\it \teod} \cite{BarbaraCosmogenic};      
\item $<1$\,nBq/kg of \coss in the {\it \teod} \cite{BarbaraCosmogenic}. This level of contamination is far below both the HPGe sensitivity and  the sensitivity achieved in \qz\ (30~$\mu$Bq/kg~\cite{Q0_2nu});
\item $< 35~\mu$Bq/kg of \coss in the \textit{CuNOSV}. 
This limit is in good agreement with the HPGe measurements performed on a few \textit{CuNOSV} samples 
($<25~\mu$Bq/kg) and also with the \coss activity resulting from the same copper\footnote{Same copper here refers to copper belonging to the same batch and having the same production history, with the only difference being a longer undergound storage period in the case of the CUORE copper.} in \qz.
\end{itemize}

\begin{table*}[htb]
\caption{Values and 90\% C.L. (95\% C.L. for HPGe measurements) upper limits on \thdt and \udt bulk contaminations of detector and cryostat materials (the various components are described in \partname~\ref{sec:exp}), as obtained in the material screening campaign. Activities are expressed in Bq/kg (1 Bq/kg = 246$\times$10$^{-9}$ g/g for \thdt and 81$\times$10$^{-9}$ g/g for \udtn). The quoted uncertainties are statistical. In the last column, the measurement technique is indicated, as described in the text. We only include the results shown in bold in the background budget evaluation, as discussed in \partname~\ref{sec:results}. }
\begin{center}
\begin{tabular}{@{}lccc@{}}

\multicolumn{4}{c}{BULK CONTAMINATIONS OF MATERIALS USED IN FAR AND NEAR REGIONS}\\
\toprule
Material 			 	 				& $^{232}$Th				 				& $^{238}$U							  	& Technique \\
         			 					    & [Bq/kg]										 & [Bq/kg]									 & \\
\noalign{\smallskip}
\midrule
\textit{TeO$_2$}	 	 	 & $<$8.4$\times$10$^{-7}$		 &  $<$6.7$\times$10$^{-7}$		 & bolometric  \\
Glue			 	 						 & $<$8.9$\times$10$^{-4}$		 &  $<$1.0$\times$10$^{-2}$		 & NAA  \\
Au bonding wires	 			 & $<$4.1$\times$10$^{-2}$		 &  $<$1.2$\times$10$^{-2}$		 & ICPMS\\
\textit{heaters}	 		 	 			 & $<$3.3$\times$10$^{-4}$		 &  $<$2.1$\times$10$^{-3}$		 & NAA \\
\textit{NTDs}		 	 & $<$4.1$\times$10$^{-3}$		 &  $<$1.2$\times$10$^{-2}$		& producer spec.\\
\textit{PEN}	 		 		 & $<$1.0$\times$10$^{-3}$		 &  $<$1.3$\times$10$^{-3}$		& NAA(Th) + HPGe(U)\\
\textit{PTFE}			 		 & $<$6.1$\times$10$^{-6}$		 &  $<$2.2$\times$10$^{-5}$		& NAA \\
\textit{CuNOSV} 		 		 			 & $<$2.0$\times$10$^{-6}$		 &  $<$6.5$\times$10$^{-5}$		 & NAA + HPGe \\
\textbf{\textit{CuOFE}} 			 				 & \textbf{$<$6.4$\times$10$^{-5}$}		 &  \textbf{$<$5.4$\times$10$^{-5}$}		 & HPGe\\
\textbf{\textit{RomanPb}}	 		 				 & \textbf{$<$4.5$\times$10$^{-5}$}	 &  \textbf{$<$4.6$\times$10$^{-5}$}		 & HPGe\\
\textbf{\textit{ModernPb}}	 		 				 & \textbf{$<$1.4$\times$10$^{-4}$}		 &  \textbf{$<$1.4$\times$10$^{-4}$}		 & NAA\\
\textbf{\textit{SI}}		 	 		 & \textbf{(11$\pm$2)$\times$10$^{-3}$}		     & \textbf{$<$2.4$\times$10$^{-3}$}		 & HPGe\\
\textbf{\textit{Rods}} 	 	 & \textbf{(4$\pm$2)$\times$10$^{-4}$ }	         & \textbf{(8$\pm$2)$\times$10$^{-4}$} 	 		& HPGe\\
\textbf{\textit{300KFlan}} & \textbf{(4.5$\pm$0.5)$\times$10$^{-3}$} 	         & \textbf{(1.6$\pm$0.5)$\times$10$^{-3}$} 	 		& HPGe\\
\noalign{\smallskip}
\bottomrule
\end{tabular}
\label{tab:bulk}
\end{center}
\end{table*}
\begin{table*}[t]
\caption{90\% C.L. upper limits on surface contamination of various CUORE detector components, as obtained in the material screening campaign. To infer the surface contamination from the measured data, different contamination depths were considered (column 1): the limits reported in the table are the ones corresponding to the contamination depth that gives the highest background contribution, as explained in the text. In the last column, the measurement technique is indicated.}
\begin{center}
\begin{tabular}{@{}lccccc@{}}
\multicolumn{6}{c}{SURFACE CONTAMINATIONS OF MATERIALS USED IN THE NEAR REGION}\\
\toprule
Material 		& Depth 	&$^{232}$Th 							& $^{238}$U 							& $^{210}$Pb 					&Technique\\
	& [$\mu$m] &[Bq/cm$^{2}$] 					&[Bq/cm$^{2}$]						&[Bq/cm$^{2}$] 									& \\
\noalign{\smallskip}
\midrule
\textit{TeO$_2$}	&0.01-10	&$<$1.9$\times$10$^{-9}$	 		&$<$8.9$\times$10$^{-9}$ 		&$<$9.8$\times$10$^{-7}$		&bolometric\\
\textit{heaters}	&0.1-10 	&$<$3.3$\times$10$^{-6}$ 	&$<$8.2$\times$10$^{-7}$	 	&$<$8.2$\times$10$^{-7}$ 	 &bolometric\\
\textit{NTDs}		&0.1-10		&$<$8.0$\times$10$^{-6}$ 	&$<$5.0$\times$10$^{-6}$		&$<$4.0$\times$10$^{-5}$ 	 &SiSB\\
\textit{PEN}		&0.1-30 	&$<$4.0$\times$10$^{-6}$ 		&$<$5.0$\times$10$^{-6}$ 		& $<$3.0$\times$10$^{-5}$		 &SiSB\\
\textit{PTFE}		&0.1-30 	&$<$1.9$\times$10$^{-8}$ 	&$<$6.8$\times$10$^{-8}$ 	& - 						 & NAA\\
\textit{CuNOSV} 	&0.1-10 	&$<$6.8$\times$10$^{-8}$		&$<$6.5$\times$10$^{-8}$ 		&$<$8.6$\times$10$^{-7}$ &bolometric\\	 
\noalign{\smallskip}
\bottomrule
\end{tabular}
\label{tab:surf}
\end{center}
\end{table*}

\subsection{\udt and \thdt bulk contamination}
\label{sec:bulkcont}
\tablename~\ref{tab:bulk} shows the \udt and \thdt bulk activities of the different CUORE materials as obtained in the radioactive assay campaign. The detection efficiency in each measurement is determined using a GEANT4-based Monte Carlo simulation that reproduces the detector geometry and distribution of contaminants in the sample.
For each isotope we report the result (or upper limit) of the  most sensitive method employed, even if more than one technique may have been used for the same material (i.e. \thdt and \udt results may come from different assay techniques for the same material).

In the case of the bulk contamination of {\it CuNOSV}, we report in \tablename~\ref{tab:bulk} the limits obtained with direct measurements and in \tablename~\ref{tab:teodbulk} the limits obtained in the \qz\ analysis (the latter are discussed in the next section). As discussed in \partname~\ref{sec:results}, the results obtained with the \qz\ detector are used for the CUORE background budget evaluation. 

The bulk contamination of the {\it NTDs} here refers to the impurity concentration certified by the manufacturer for undoped wafers. The doping is done in a nuclear reactor producing a large number of short-lived radioactive isotopes. The hypothesis that long-lived isotopes potentially dangerous for a \BBz bolometric experiment could also be produced during reactor exposure was investigated in~\cite{articoloNTD} and rejected. After the doping process, an ohmic contact is created on the Ge surface. This operation could cause contamination of the thermistor, that is analyzed as a surface contribution. 
It is anyway worth noting that 
signals originating from nuclear decays occurring in the thermistor volume are deformed in their shape, and can therefore be efficiently rejected by the standard pulse shape cuts applied by the analysis process.

\subsection{\udtn, \thdt and \pbdd surface contamination}
\label{sec:surfchains}
In \tablename~\ref{tab:surf}, we report the most sensitive upper limits obtained for the surface activities of the materials used in the near region.

As it was done for the bulk contamination, a Monte Carlo simulation was used to determine, in each measurement, the detection efficiency  of the surface contamination. While for bulk contamination the free parameter is only the bulk activity of the sample (always considered to be uniformly distributed in its volume), in the case of surface contamination the impurity distribution is described according to Eq.~\ref{eq:exp} using two parameters. In all the measurements discussed in this section, experimental data are not enough to measure both $\lambda$ and $\rho_0$. Therefore we proceed by evaluating, for each $\lambda$,  the contaminant density $\rho_0$ compatible with the experimental data. The integral of $\rho$ over the contaminated volume, divided by the surface of the sample yields the impurity concentration measured in Bq/cm$^2$. After having evaluated the surface impurity concentration for a wide range of depths ($\lambda$), we chose the one producing the highest background contribution, and quoted this limit in Table\,\ref{tab:surf}.  The minimum depth considered in this analysis is 0.001 $\mu$m for {\it \teod} and 0.1 $\mu$m for any other material. Indeed, these are the most shallow distributions whose effects can be experimentally identified. The maximum depth is \ca 10 $\mu$m for  {\it \teod} and {\it CuNOSV}  (the range of 5~MeV $\alpha$ particles in these materials is \ca 10\,$\mu$m and \ca 15\,$\mu$m, respectively), and 30\,$\mu$m for {\it PTFE} (where the range of 5~MeV $\alpha$ particles is \ca 23\,$\mu$m). Larger depths are almost indistinguishable from bulk contaminations. 

While {\it NTDs} and {\it PEN} were measured through $\alpha$ spectroscopy with SiSB diodes, in the case of the {\it heaters} the small size of the sample required a more sensitive technique. The $\alpha$ spectrometer used in this case was an array of two \ciccio \teod bolometers operated in the cryogenic facility of Hall C at LNGS. A matrix of {\it heaters}, 5$\times$5~cm$^2$ in total area, was oriented towards the crystals and the $\alpha$ induced background was analyzed with the technique  illustrated in~\cite{rad2006}.
In the case of \udt and \thdt impurities in  {\it PTFE}, the best upper limits are obtained with the NAA technique (analyzing the results as if all measured contaminants were contained in a surface layer). Giving information only on the progenitor, such technique is insensitive to \pbdd contamination.
A bolometric detector array (TTT, for Three Tower Test) was used to compare different copper surface treatments and to analyze the {\it CuNOSV} contamination level. The TTT detector \cite{TTT} consisted of three small towers, each with 12 \teod detectors and enclosed inside its own copper box. The limits reported in \tablename~\ref{tab:surf} are those obtained with the TECM cleaning, wich showed the lowest background level and was therefore chosen as the baseline cleaning protocol for all the {\it CuNOSV} pieces of the CUORE detector.

\begin{table}[t]
\caption{Values and 90\% C.L. upper limits on bulk contaminations in the \teod crystals and the \textit{Holder} based on the background model of \qz\ \cite{Q0_2nu}. Contaminants are identified as follows: when the progenitor is indicated all the chain is assumed in secular equilibrium, in all other cases single isotopes or sub-chains of \thdt and \udt are considered.  Each contaminant is assigned a unique index for future reference.}
\begin{center}
	\begin{tabular}{@{}llcr@{}}
\multicolumn{4}{c}{BULK CONTAMINATION of \textit{TeO$_2$} and \textit{Holder} }\\
\toprule

Material				&Source				&Index				& Activity [Bq/kg] \\
\noalign{\smallskip}
\midrule
\noalign{\smallskip}
\multirow{7}{*}{\textit{\teod}}	&$^{210}$Po				&1					&(2.39$\pm$0.11)$\times$10$^{-6}$\\
					    &$^{210}$Pb				&2					&(1.37$\pm$0.19)$\times$10$^{-6}$\\
						&$^{232}$Th only			&3					&(7$\pm$3)$\times$10$^{-8}$\\
						&$^{228}$Ra-$^{208}$Pb		&4		&$< $\,3.5$\times$10$^{-8}$\\
						&$^{238}$U--$^{230}$Th			&5					&$<$\,7.5$\times$10$^{-9}$\\
						&$^{230}$Th only			&6					&(2.8$\pm$0.3)$\times$10$^{-7}$\\
						&$^{226}$Ra--$^{210}$Pb		&7			&$< $\,7$\times$10$^{-9}$\\
\midrule
\multirow{2}{*}{\textit{Holder}}	&\thdt				&18					&$<$\,2.1$\times$10$^{-6}$ \\
						&\udt				&19					&$<$\,1.2$\times$10$^{-5}$ \\
\noalign{\smallskip}
\bottomrule
\end{tabular}
\label{tab:teodbulk}
\end{center}
\end{table}

\begin{table}[t]
\caption{90\% C.L. upper limits and values for surface contamination of the \teod crystals and the \textit{Holder} based on the background model of \qz\ \cite{Q0_2nu}. Contaminants are identified as follows: when the progenitor is indicated all the chain is assumed in secular equilibrium, in all other cases single isotopes or sub-chains of \thdt and \udt are considered. Each contaminant is assigned a unique index for future reference.}
\begin{center}
\begin{tabular}{@{}llcr@{}}
\multicolumn{4}{c}{SURFACE CONTAMINATION of \textit{\teodn} and \textit{Holder}}\\
\toprule
Material				& Contamination				&Index		    & Activity [Bq/cm$^{2}$]\\
\midrule
\multirow{10}{*}{\textit{\teod}} &$^{232}$Th only .01$\mu$m		&8				&(3$\pm$1)$\times$10$^{-10}$\\
						&$^{228}$Ra-$^{208}$Pb .01$\mu$m		&9				&(2.32$\pm$0.12)$\times$10$^{-9}$\\
						&$^{238}$U-$^{230}$Th .01$\mu$m		&10				&(2.07$\pm$0.11)$\times$10$^{-9}$\\
						&$^{230}$Th only .01$\mu$m		&11				&(1.15$\pm$0.14)$\times$10$^{-9}$\\
						&$^{226}$Ra-$^{210}$Pb .01$\mu$m		&12				&(3.14$\pm$0.10)$\times$10$^{-9}$\\
						&$^{210}$Pb .001$\mu$m			&13				&(6.02$\pm$0.08)$\times$10$^{-8}$\\
						&$^{210}$Pb 1$\mu$m				&14				&(8.6$\pm$0.8)$\times$10$^{-9}$\\
						&$^{210}$Pb 10$\mu$m				&15				&$<$\,2.7$\times$10$^{-9}$\\
						&$^{232}$Th 10$\mu$m				&16				&(7.8$\pm$1.4)$\times$10$^{-10}$\\
						&$^{238}$U 10$\mu$m				&17				&$<$\,3.3 $\times$10$^{-11}$\\
\midrule
\multirow{5}{*}{\textit{Holder}}
						&$^{210}$Pb .01$\mu$m			&20				&(2.9$\pm$0.4)$\times$10$^{-8}$\\
						&$^{210}$Pb .1$\mu$m				&21				&(4.3$\pm$0.5)$\times$10$^{-8}$\\
						&$^{210}$Pb 10$\mu$m				&22				&$<$1.9 $\times$10$^{-8}$\\
						&$^{232}$Th 10$\mu$m				&23				&(5.0$\pm$1.7)$\times$10$^{-9}$\\
						&$^{238}$U 10$\mu$m				&24				&(1.38$\pm$0.16)$\times$10$^{-8}$\\

\noalign{\smallskip}
\bottomrule
\end{tabular}
\label{tab:teodsurf}
\end{center}
\end{table}

\subsection{Contamination from \qz\ analysis}\label{sec:surfcont}

In~\cite{Q0_2nu}, the sources contributing to the \qz\ event rate were reconstructed by fitting a number of Monte Carlo simulations to the measured spectra. The fit was performed with a Bayesian approach which allows to exploit any previous knowledge on material contamination by defining \textit{priors} on source activities. For many sources, a sensitivity on contaminant concentration better than that achieved with standard techniques was obtained, along with a more efficient disentanglement of contamination species and a detailed study of secular equilibrium violations in radioactive chains. 

\begin{figure} [t]
\begin{center}
\includegraphics[width=0.5\textwidth]{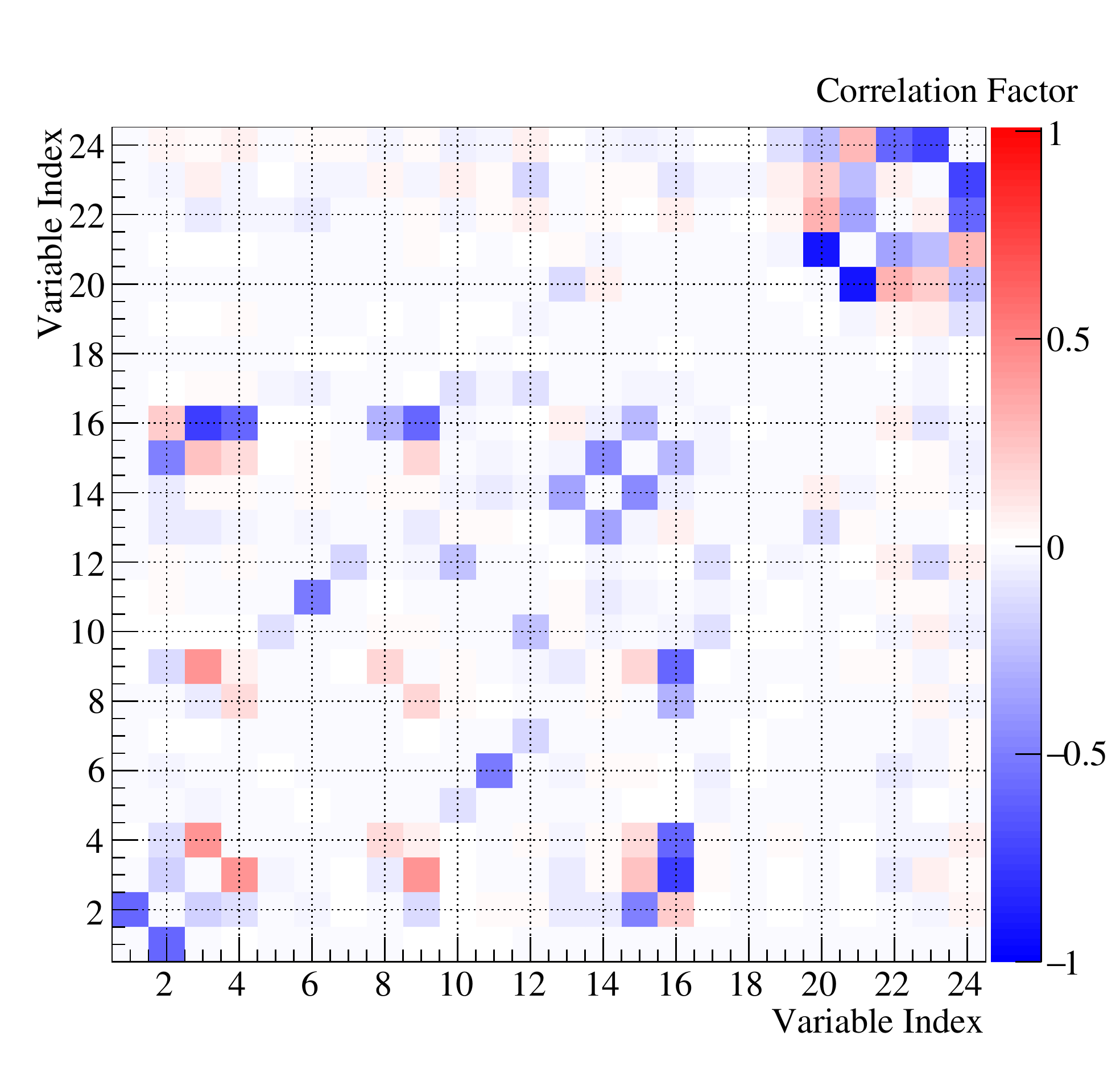}
\caption{Correlation matrix among the activities evaluated in the \qz\ background reconstruction. The list of sources is reported in Tables~\ref{tab:teodbulk} and ~\ref{tab:teodsurf}
with their Index numbering.}
\label{fig:corrmatrix}
\end{center}
\end{figure}

\begin{figure} [hbt]
\begin{center}
\includegraphics[width=0.5\textwidth]{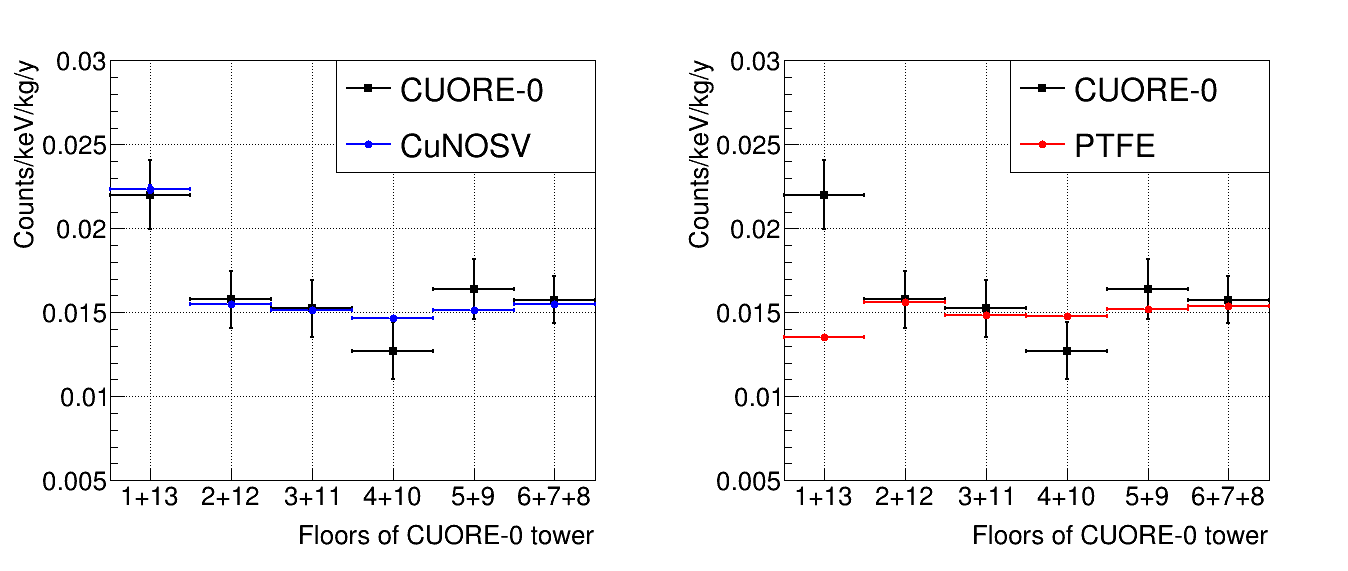}
\caption{\qz\ plane-by-plane experimental rate in the energy region between 2.7\,MeV and 3.9\,MeV (black points) compared to the plane-by-plane Monte Carlo simulated rate of a \pbdd\ contamination in the NOSV copper detector structure (left plot) and to the plane-by-plane Monte Carlo simulated rate of a \pbdd\ contamination in the crystal PTFE supports (right plot). To increase the statistical significance, the data from detector planes in symmetric positions around the middle  of the tower (floor 7) have been grouped.}
\label{fig:floor-rate}
\end{center}
\end{figure}

\tablename~\ref{tab:teodbulk} and \tablename~\ref{tab:teodsurf} summarize the results obtained for the impurity concentrations in materials that belong to the same production batches of CUORE, i.e. the \teodn\ crystals and the NOSV copper (that in \qz\ analysis was taken as representative of the whole detector holder structure ---see later).
For the bulk contamination levels, the sensitivity improvements obtained by the \qz\ analysis are minor (compare the limits reported for \textit{\teodn} in \tablename~\ref{tab:bulk} and \ref{tab:teodbulk} or those reported for \textit{CuNOSV} in \tablename~\ref{tab:bulk} and for \textit{Holder} in \tablename~\ref{tab:teodbulk}). On the contrary, for 
surface contaminations the improvements are significant: the \qz\ analysis allowed 
to derive a model for the impurity density profiles meanwhile achieving a high sensitivity in the determination of their activity.
Moreover, since the procedure adopted in \qz\ background reconstruction was to simultaneously fit all the simulated spectra to the measured one, it was possible  to find the best evaluation for each contaminant concentration as well as the correlation factor among different sources (see Fig.~\ref{fig:corrmatrix}).
\\A separate discussion applies to the \textit{small parts} used to build the \qz\ detector --- i.e. NTD thermistors, PEN cables, Si heaters, Au bonding wires, glue and PTFE supports --- also belonging to the same production batch of CUORE. In fact, in \qz\ analysis these small mass/surface components exhibit spectra that are completely degenerate with those of the NOSV copper detector structure; therefore only the latter was considered in the model under the name of \textit{Holder} (see \tablename~\ref{tab:teodbulk} and \tablename~\ref{tab:teodsurf}), chosen to underline that the resulting activities also include the \textit{small parts} contribution. This contribution 
is expected to be negligible in the \qz\ background, as confirmed by Fig.~\ref{fig:floor-rate}. 
In these plots the experimental rate of the different planes of the \qz\ detector in the energy region between 2.7\,MeV and 3.9\,MeV (i.e. the interval dominated by degraded  $\alpha$ contributions from surface contaminants) is compared to the rate expected from a \pbdd\ contamination in the NOSV copper detector structure or in the PTFE supports (chosen as representative of all the \textit{small parts}, because of their bigger mass/surface) ---similar plots are obtained for \thdt\ and \udt\ contaminants. 
The degeneracy between these two sources can be broken by examining the detector plane dependence of the counting rate. This is illustrated in Fig.~\ref{fig:floor-rate}. The detectors in the upper and lower planes of the \qz\ tower (1+13 floors in Fig.~\ref{fig:floor-rate}) face a larger copper surface of the 10\,mK shield, while seeing a slightly smaller surface of PTFE and other small parts compared to the middle floors of \qz. The \qz\ data shown in Fig.~\ref{fig:floor-rate} suggest that the contribution to the overall background in the 2.7-3.9 MeV region from PTFE supports is compatible with zero. 
The resulting  PTFE supports surface contaminations are reported in \tablename~\ref{tab:ptfesurf} (to be compared to the corresponding ones in \tablename~\ref{tab:surf}). 
\begin{table}[t]
\caption{90\% C.L. upper limits for surface contamination of the PTFE supports based on the background model of \qz\ \cite{Q0_2nu} with the additional information coming from Fig.~\ref{fig:floor-rate}.}
\begin{center}
\begin{tabular}{@{}llr@{}}
\multicolumn{3}{c}{SURFACE CONTAMINATION of \textit{PTFE}}\\
\toprule
Material				& Contamination					    & Activity [Bq/cm$^{2}$]\\
\midrule
\multirow{3}{*}{\textit{PTFE}}
&$^{210}$Pb 10$\mu$m				&$<$7.6 $\times$10$^{-8}$\\
&$^{232}$Th 10$\mu$m				&$<$1.5 $\times$10$^{-8}$\\
&$^{238}$U 10$\mu$m					&$<$1.9 $\times$10$^{-8}$\\
\noalign{\smallskip}
\bottomrule
\end{tabular}
\label{tab:ptfesurf}
\end{center}
\end{table}
%

\section{Background budget}\label{sec:results}
\begin{figure*}[t]
\centering
\subfigure[]
{\includegraphics[width=.45\linewidth]{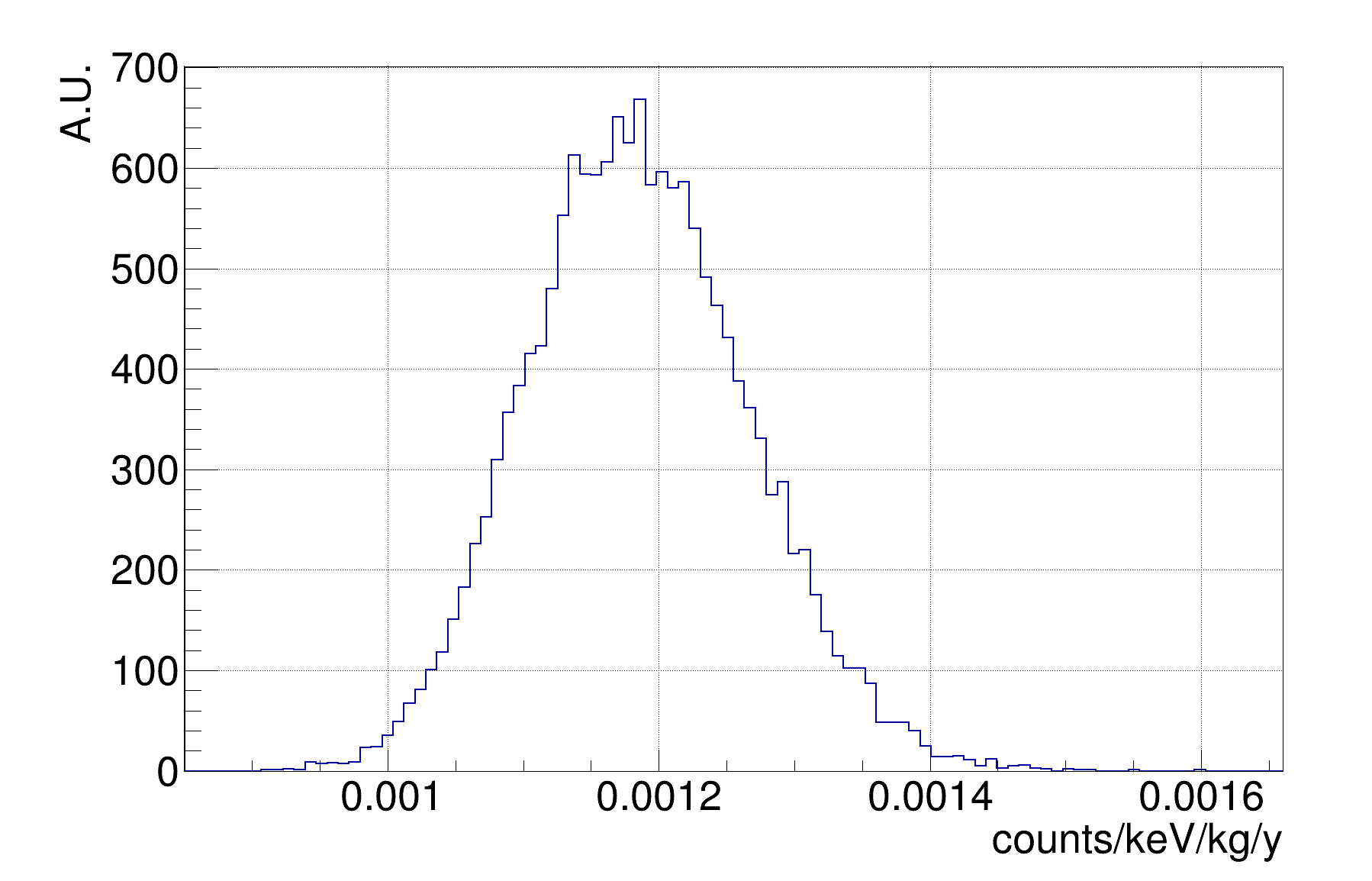}}
\hspace{5mm}
\subfigure[]
{ \includegraphics[width=.45\linewidth]{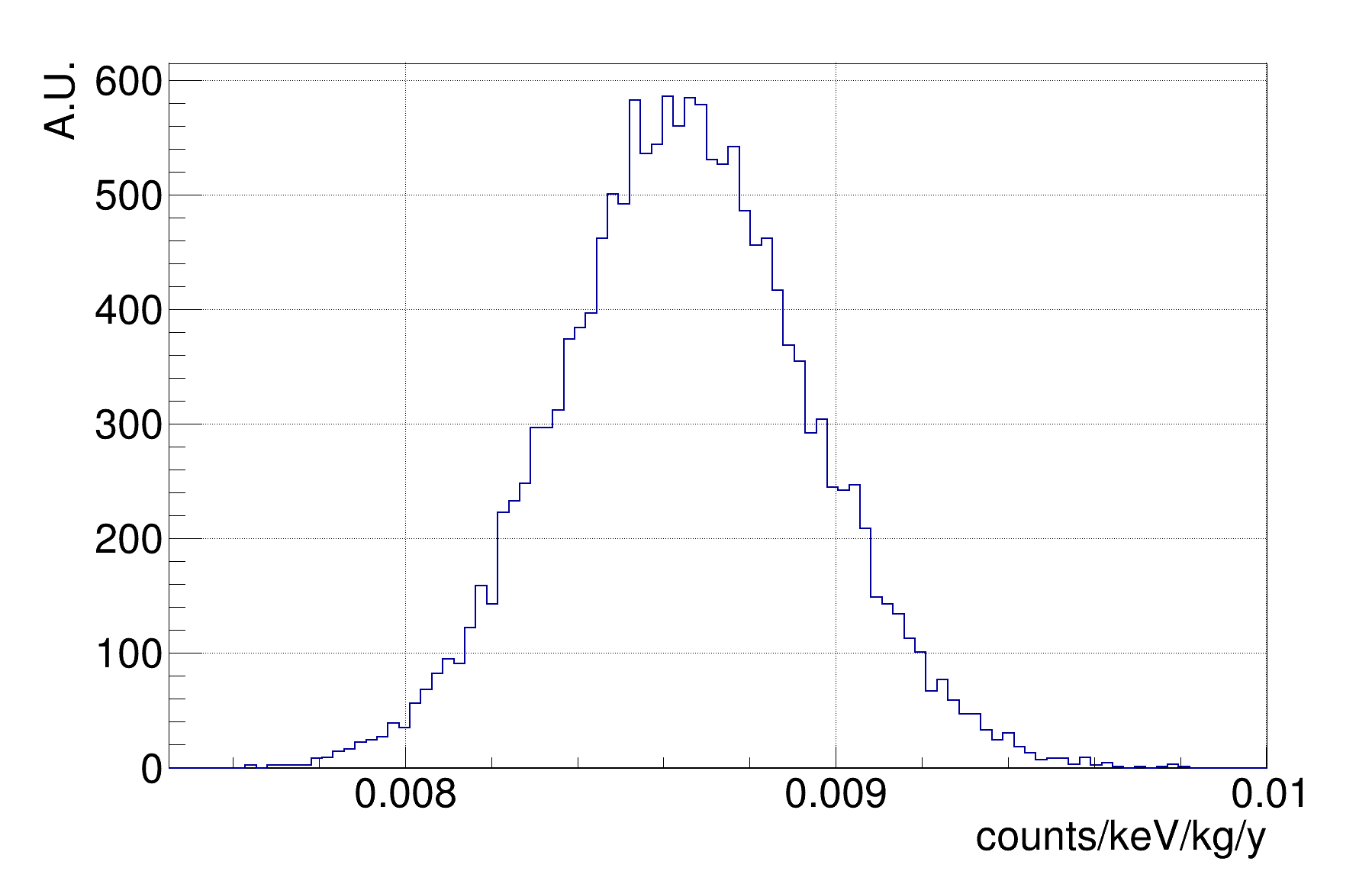}}
\caption{Global pdf's, accounting for internal correlations, for the BI contribution of  (a) \textit{\teod} bulk and surface impurities and  (b) \textit{CuNOSV} bulk and surface impurities. The mean of each pdf is taken as the estimation of the global contribution to the CUORE background budget BI$_i$. 
}
\label{fig:correlati}
\end{figure*}

In this section we discuss contributions from material radioactivity to the CUORE background.  From now on, the event rate in the ROI will be denoted as the Background Index, BI, and its estimation will include also the instrumental efficiency (discussed later), in order to properly represent the experimental event rate expected in CUORE.

Following the discussion of the previous sections, the sources that give sizable contribution to the BI are:
\begin{enumerate} 
\item (near region) \udtn, \thdt and their progenies in \textit{\teodn} and \textit{Holder}  ---the latter as representative of both the \textit{CuNOSV} and the \textit{small parts}. For both bulk and surface activities of these sources we use \qz\ results as reported in Tables~\ref{tab:teodbulk} and~\ref{tab:teodsurf} along with the respective correlation factors;
\item (near region) cosmogenically activated isotopes in \textit{\teodn} and \textit{CuNOSV}, using the activities discussed in \partname~\ref{sec:bulkcont};
\item (far region) \udt and \thdt in \textit{CuOFE}, \textit{RomanPb}, \textit{ModernPb}, \textit{SI}, \textit{Rods}, \textit{300KFlan} (refer to Table~\ref{tab:masses}). Activities are those reported in Table~\ref{tab:bulk}.
\end{enumerate}

The BI is evaluated as follows. For each source $i$ (i.e. a definite detector element contaminated by a specific contaminant) we run a MC simulation and project the simulated spectra for the 988 bolometers. As we plan to do with the experimental data, we enforce an anticoincidence cut (i.e. we reject multi-hit events). The 988 spectra are finally summed and the source efficiency ($\epsilon_i^{MC}$) is evaluated. $\epsilon_i^{MC}$ is defined as the probability for the source $i$ to produce an event in the ROI of the sum spectrum; it is therefore given by the ratio between the number of \textit{single-hit} events in the ROI ($N^{i}_{ROI}$) and the number of simulated decays ($N^{i}_{decays}$)\footnote{In the case of \coss cosmogenic activation in \textit{CuNOSV}, the 2505 keV peak contribution is subtracted.}:

\begin{equation}\label{eq:MCeffi}
  \epsilon_i^{MC}=N^{i}_{ROI}/N^{i}_{decays}.
\end{equation}

In the case of sources with uncorrelated activities (points 2 and 3 in the previous list), individual contributions to the BI are given by: 


\begin{equation}\label{eq:BI}
	\textrm{BI}_i =\frac{A_i \times (\epsilon_i^{MC} \times \epsilon^{instr})}{\Delta  \times M}
\end{equation}

\noindent where $A_i$ is the activity of the source $i$, $\Delta$ is the ROI width (100 keV) and $M$ is the \teod mass. 
The factor $\epsilon_i^{MC} \times \epsilon^{instr}$ measures the probability of observing a contribution to the measured ROI event rate caused by source $i$. It is a product of conditional probabilities~\cite{Q0-analysis}: the probability for the source $i$ to produce an event in the ROI ($\epsilon_i^{MC}$), the probability that this event is triggered and properly reconstructed ($\epsilon^{tr}$),
the probability that this event is not accidentally in coincidence with an unrelated event in a different bolometer ($\epsilon^{acc}$) and the probability that it passes the pulse shape cuts ($\epsilon^{PSA}$). These cuts are used to remove events that are either non-signal-like or are in some way not handled well by the data processing software~\cite{Q0-analysis}.

$\epsilon^{tr}$ and $\epsilon^{PSA}$ depend on data acquisition and analysis, and we expect them to be similar to CUORE-0 ones~\cite{Q0-analysis}, $\epsilon^{tr}$=\,98.5\% and $\epsilon^{PSA}$=\,93.7\%. $\epsilon^{acc}$ depends on the probability of accidental coincidences and is evaluated to be 99\% (as expected for a $\pm$ 5~ms coincidence window and an average signal rate of 1~mHz/bolometer).
The product of $\epsilon^{acc}$, $\epsilon^{tr}$ and $\epsilon^{PSA}$ is source independent (we assume here all efficiencies to be energy independent), and we will refer to it as instrumental efficiency, $\epsilon^{instr}$=\,91.4\%.

The BIs computed following this procedure have a statistical uncertainty that is obtained propagating the statistical uncertainty of A$_i$ and a systematic error that derives both from A$_i$ (material activity measurements have systematics of the order of $\pm$5\%) and from $\epsilon_i^{MC}$ ($\pm$5\%, due to approximations in the description of geometry and detector response to particle interaction).
\begin{figure*}[t]
\begin{center}
\includegraphics[width=.99
\linewidth]{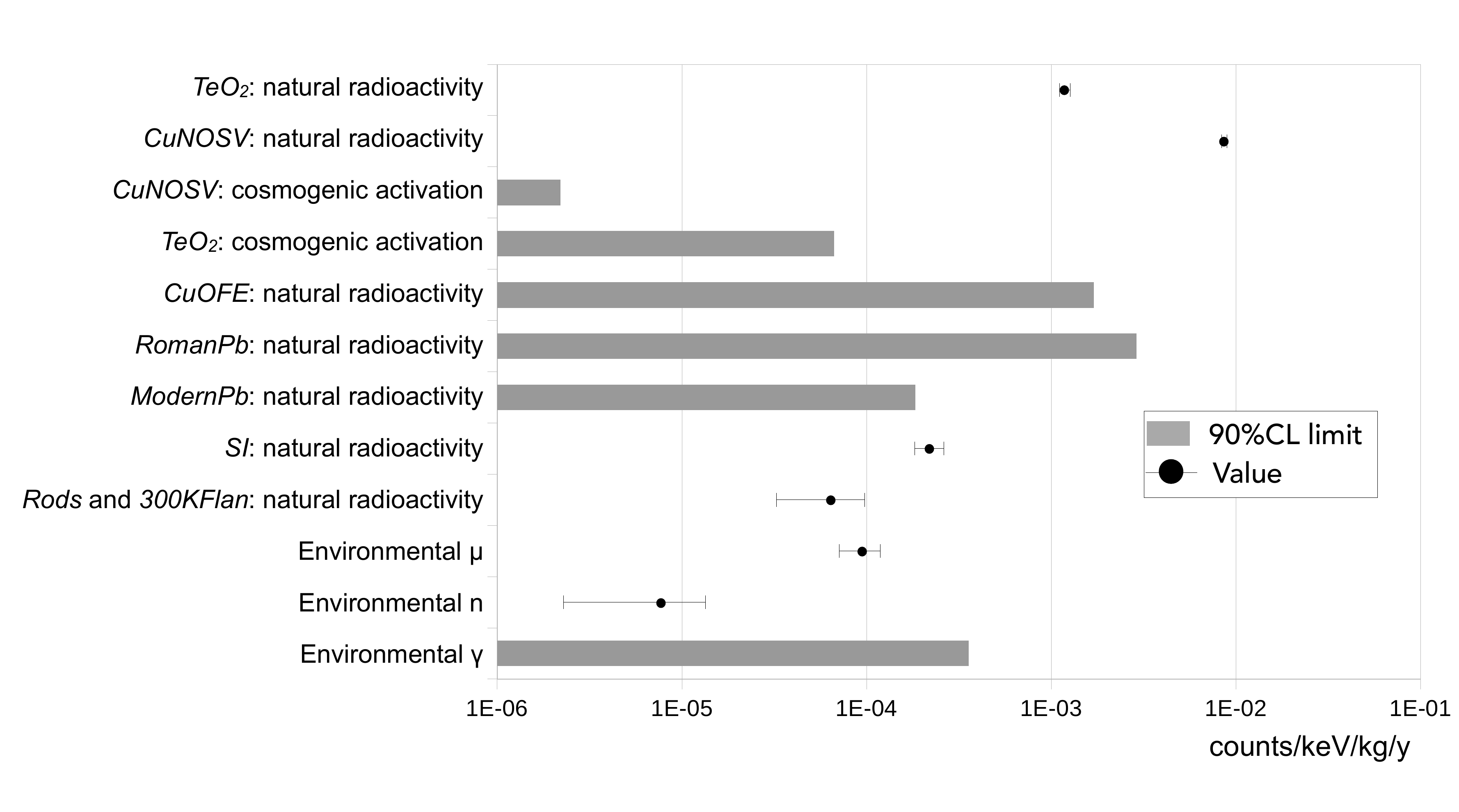}
\end{center}
\caption{Histogram representing the main BI expected for the various components of CUORE.  The grey bars indicate 90\%~C.L. upper limits while the dots (with 1$\sigma$ uncertainties) indicate derived values. Only statistical uncertainties are indicated.}
\label{fig:istoBB}
\end{figure*}

\begin{table*}[t]
\caption{Main background indexes expected for the various components of CUORE (limits are 90\% C.L.).}
\begin{center}
\begin{tabular}{@{}lll@{}}
\noalign{\smallskip}
\toprule
\multirow{2}{*}{Region}		&\multirow{2}{*}{Source}					& Background Index\\
							&											& [counts/keV/kg/y]\\
\noalign{\smallskip}
\midrule
\noalign{\smallskip}
\multirow{3}{*}{External }	&Environmental \textit{$\gamma$}				&$<$3.6$\times$10$^{-4}$\\
							&Environmental \textit{n}						&(8$\pm$6)$\times$10$^{-6}$\\	
							&Environmental $\mu$						&(1.0$\pm$0.2)$\times$10$^{-4}$\\
\noalign{\smallskip}
\midrule
\multirow{5}{*}{Far}		&\textit{Rods} and \textit{300KFlan}: natural radioactivity	&(7$\pm$3)$\times$10$^{-5}$\\
							&\textit{SI}: natural radioactivity						&(2.2$\pm$0.4)$\times$10$^{-4}$\\	
							&\textit{ModernPb}: natural radioactivity					&$<$1.8$\times$10$^{-4}$\\
							&\textit{RomanPb}: natural radioactivity					&$<$2.9$\times$10$^{-3}$\\
							&\textit{CuOFE}: natural radioactivity 					&$<$1.7$\times$10$^{-3}$\\
\noalign{\smallskip}
\midrule
\multirow{4}{*}{Near}		&\textit{\teod}: cosmogenic activation					&$<$6.7$\times$10$^{-5}$\\
							&\textit{CuNOSV}: cosmogenic activation 					&$<$2.2$\times$10$^{-6}$\\
							&\textit{CuNOSV}: natural radioactivity					&(8.7$\pm$0.3)$\times$10$^{-3}$\\
							&\textit{\teod}: natural radioactivity					&(1.2$\pm$0.1)$\times$10$^{-3}$\\
\noalign{\smallskip}
\bottomrule
\end{tabular}
\label{tab:summary}
\end{center}
\end{table*}

In the case of the activities extrapolated from \qz\ analysis (sources at point 1 in the previous list), the determination of the background index follows a slightly different procedure. \figurename~\ref{fig:corrmatrix} shows that we have two groups of sources,\textit{ \teodn} contaminants and \textit{Holder} contaminants, with strong correlations within each group.
Thus we define two cumulative Background Indexes, BI$_{ TeO_2}$ and BI$_{Holder}$, 
and we associate to both of them a probability density function (pdf), obtained as follows.
The joint pdf describing contaminant activities, as derived by the CUORE-0 background fit reconstruction, is sampled. For each sampled point, the activity of the $i^{th}$ source is weighted by its respective CUORE efficiency $\epsilon_{i}^{MC} \times \epsilon^{instr}$ and the sum of the activities induced in the ROI by sources belonging to the same group is calculated, thus taking into account the correlations.
The pdf's associated with BI$_{TeO_2}$ and BI$_{Holder}$ are shown in \figurename~\ref{fig:correlati}. In both cases, bulk contamination gives a minor contribution (1--2 orders of magnitude lower) with respect to surface one, which represents therefore the dominant source. The mean of each pdf is taken as the estimation of the BI and the RMS is used to compute its statistical uncertainty.\\
In the adopted procedure, BI$_{Holder}$ is computed under the hypothesis of neglecting the \textit{small parts} contribution to the ROI rate. This assumption, though completely acceptable in the case of \qz\ analysis (see discussion in Section~\ref{sec:surfcont}), may not be thoroughly valid in CUORE because of the different scaling of the geometrical efficiencies of \textit{CuNOSV} and \textit{small parts} going from \qz\ to CUORE. As a matter of fact, the various towers of the CUORE detector are facing the same amount of \textit{small parts} surface but  different amounts of \textit{CuNOSV} surface, depending on their position in the array with respect to the \textit{CuNOSV} thermal shield of the cryostat. 
Though \textit{small parts} contribution is compatible with zero in the case of \qz\ model, their introduction in  CUORE model adds an additional +15\% systematic error to BI$_{Holder}$. Therefore, the final systematic errors are  $\pm$10\% for BI$_{TeO_2}$ and $_{-10\%}^{+25\%}$ for BI$_{Holder}$. 


\tablename~\ref{tab:summary} and \figurename~\ref{fig:istoBB} summarize the final results: in both we report also the contribution expected from external sources (``External'') as derived in~\cite{CuoreExternal}.

The total projected BI in the ROI of CUORE is equal to [1.02$\pm$0.03(stat)$^{+0.23}_{-0.10}$(syst)]$\times$10$^{-2}$\,counts/keV/kg/y. This number is obtained by summing all the contributions incompatible with zero listed in \tablename~\ref{tab:summary} (i.e. we exclude sources for which we have an upper limit). The BI is by far dominated by the \textit{Holder} contribution, mainly ascribed to degraded alpha particles from surface contaminants (see~\cite{Q0_2nu}). 

The CUORE predicted BI can be compared to the  BIs measured by precursor experiments (in units of counts/keV/kg/y): 
\begin{itemize}
\item[$\bullet$] 
$0.058 \pm 0.004$\,(stat.)\,$\pm$\,0.002\,(syst.) ($\epsilon^{instr}$= 92.0\%) in CUORE-0; 
\item[$\bullet$] 
$0.153 \pm 0.006$\  ($\epsilon^{instr}$= 94.7\%) in \hbox{Cuoricino}.
\end{itemize}
Finally, we report in Table\,~\ref{tab:layer} the BIs in different subgroups of detectors (``layers''), grouped by the number of crystal sides directly facing the \textit{CuNOSV} thermal shield. Layer-0 contains those crystals with no sides facing the \textit{CuNOSV} shield (i.e. \teod crystal fully shielded by other detectors), Layer-1 includes crystals with only one side facing the \textit{CuNOSV} shield, and so on. This grouping reflects the fact that contaminations on the \textit{CuNOSV} surface are the most significant contribution to the background in the ROI. As a result, we expect different layers to have different projected sensitivities to the \BBz half-life of $^{130}$Te, resulting in a global improvement in the final CUORE sensitivity.

\begin{table}[thb]
\caption{Expected background indexes (statistical errors only) for the different layers of the CUORE tower array. The number of crystals in each layer is reported.}
\begin{center}
\begin{tabular}{@{}ccc@{}}
\noalign{\smallskip}
\toprule
\multirow{2}{*}{Layer}	&\multirow{2}{*}{Number of crystals}	&Background Index\\
    					&							&[counts/keV/kg/y]\\
\noalign{\smallskip}
\midrule
\noalign{\smallskip}
0						&528						&(0.82$\pm$0.02)$\times$10$^{-2}$\\
1						&272						&(1.17$\pm$0.04)$\times$10$^{-2}$\\
2						&164						&(1.36$\pm$0.04)$\times$10$^{-2}$\\
3						&24							&(1.78$\pm$0.07)$\times$10$^{-2}$\\
\noalign{\smallskip}
\bottomrule
\end{tabular}
\label{tab:layer}
\end{center}
\end{table}

\begin{figure*}[h!]
\begin{center}
\includegraphics[width=1\linewidth]{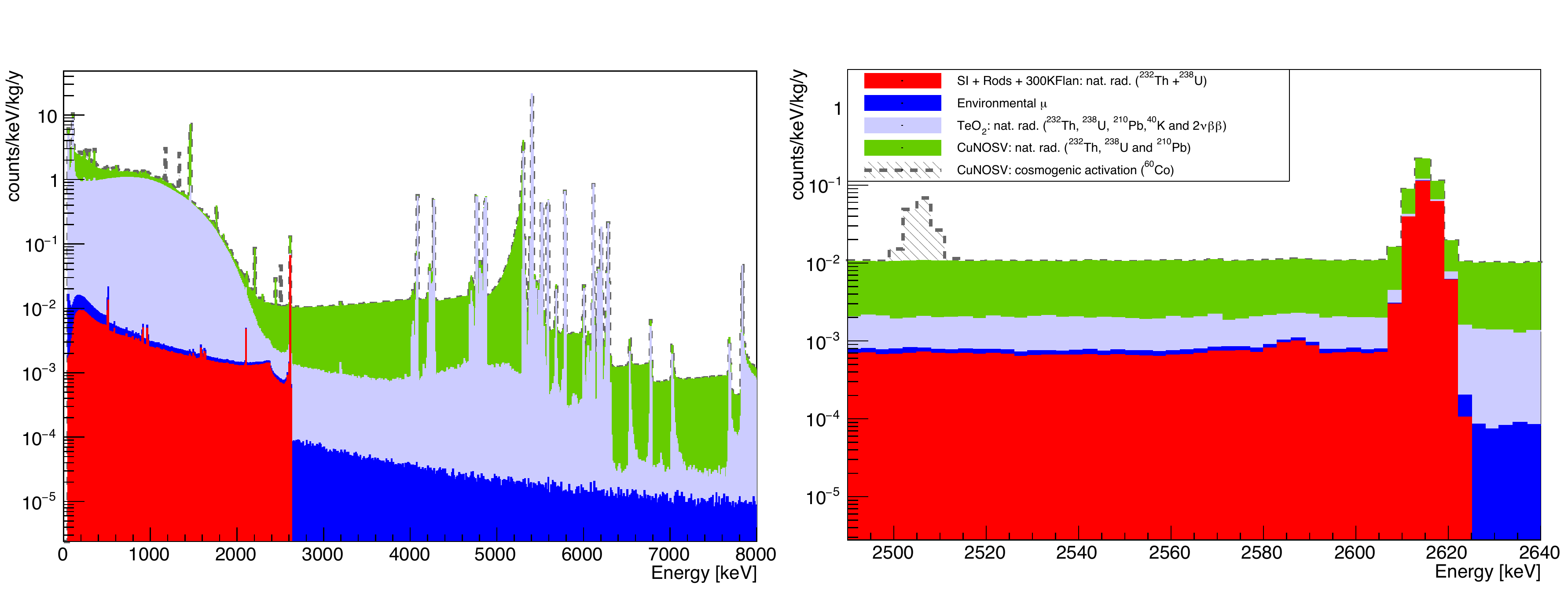}
\caption{Anticoincidence spectrum resulting from the sources analyzed in this paper (histograms corresponding to the different sources are stacked one over the other, the dotted contour shows the total predicted spectrum). Only sources with activities incompatible with zero are considered, with the exception of the \coss contamination in \textit{CuNOSV} (reported at its 90\% C.L. upper limit). Due to their expected large contribution to the shape of the spectrum, the \kq in \textit{\teodn} and in \textit{CuNOSV} as well as the \tect \BBd decay contributions are also included. The region around the ROI (i.e. the interval 2470-2570~keV) is shown in the right panel. Two peaks are visible, they are: the \coss line at 2505~keV line and the \tld line at 2615~keV (see text for more details).}
\label{fig:spettro}
\end{center}
\end{figure*}


%

The sources analyzed in this study are expected to be the major contributors to CUORE background index. Therefore, the results of this work can be used for a tentative prediction of the CUORE spectrum over a wide energy range.
Fig.~\ref{fig:spettro} shows the anticoincidence spectrum recorded by the 988 bolometer array predicted by this analysis.  Among the contaminants analyzed in this paper, we summed only those that have activities incompatible with zero (i.e. those used for the BI evaluation), with the exception of  \coss in the \textit{CuNOSV}, which is fixed at its 90\% C.L. upper limit.
Indeed, even if it does not contribute to the BI, it is likely that the 2505~keV peak will appear in the ROI as it did for CUORE-0, although its activity in CUORE has a large uncertainty.
Three additional sources are considered because of their high activity: \tect \BBd decay and \kq in \textit{\teodn} and {\it CuNOSV} bulk, all fixed at the \qz\ results~\cite{Q0_2nu}. Due to the lack of information on \kq activities of other materials, the 1460~keV \kq peak could be underestimated.

The right panel of Fig.~\ref{fig:spettro} shows the predicted spectrum in the region around the ROI. 
Two peaks are visible, one from \coss at 2505~keV (discussed earlier) and one from \tld at 2615~keV.

\section{Conclusion}
In this paper the different sources contributing to the CUORE background figure of merit are analyzed in detail. The study exploit the available data on bulk and surface contamination of CUORE construction materials and is based on the final design of the experiment.
 
All possible background sources are investigated by means of accurate dedicated measurements, and using the most sensitive available techniques with realistic experimental setup. 
A specially developed simulation code allows us to calculate the effects of the various background sources and to compare them with the required experimental sensitivity.

A dedicated R\&D program (for material selection or improvement of the measurement sensitivity) was initiated each time an experimental result was not consistent with the corresponding design sensitivity. 

The equivalence of the CUORE and the \qz\ detector tower structure and materials allowed us to fully exploit the results of the \qz\ background model reference fit \cite{Q0_2nu} as input for the activity of all the common sources (mainly \textit{\teodn} and \textit{CuNOSV} bulk and surface contaminations).

Results of the background model developed in this work are reported in \tablename~\ref{tab:summary}. None of the bulk contaminations appear worrisome for the CUORE target background rate of 10$^{-2}$ counts/keV/kg/y. Despite being more serious, surface contaminations of detector components (the detectors themselves and the materials directly surrounding them), are nevertheless within the CUORE goals for the background counts. The final estimate for the BI is [1.02$\pm$0.03(stat)$^{+0.23}_{-0.10}$(syst)]$\times$10$^{-2}$\,counts/keV/kg/y.

In conclusion, as shown in \figurename~\ref{fig:istoBB}, the work and processes involving material selection and surface cleaning were successful in yielding a result consistent with the background budget goal of the CUORE experiment.

%


\section{Acknowledgments}
The CUORE Collaboration thanks the directors and staff of the
Laboratori Nazionali del Gran Sasso and the technical staff of our
laboratories. This work was supported by the Istituto Nazionale di
Fisica Nucleare (INFN); the National Science
Foundation under Grant Nos. NSF-PHY-0605119, NSF-PHY-0500337,
NSF-PHY-0855314, NSF-PHY-0902171, NSF-PHY-0969852, NSF-PHY-1307204, NSF-PHY-1314881, NSF-PHY-1401832, and NSF-PHY-1404205; the Alfred
P. Sloan Foundation; the University of Wisconsin Foundation; and Yale
University. This material is also based upon work supported  
by the US Department of Energy (DOE) Office of Science under Contract Nos. DE-AC02-05CH11231,
DE-AC52-07NA27344, and DE-SC0012654; and by the DOE Office of Science, Office of Nuclear Physics under Contract Nos. DE-FG02-08ER41551 and DE-FG03-00ER41138.
This research used resources of the National Energy Research Scientific Computing Center (NERSC).

\bibliographystyle{spphys}       
\bibliography{bib}   

\end{document}